\documentclass[aip, jap, reprint, nobalancelastpage]{revtex4-1}

\usepackage{graphicx}
\usepackage{rotating}
\usepackage{amssymb}
\usepackage{amsmath}
\usepackage{mathptmx}
\usepackage{comment}
\usepackage[usenames, dvipsnames, svgnames, table]{xcolor}
\usepackage{url}
\usepackage{siunitx}
\usepackage{booktabs}

\newcommand{\unit}[1]{~\mathrm{#1}}
\newcommand{\fcw}{148}  % The continuous-wave frequency in GHz.
\newcommand{\beginsupplement}{
        \setcounter{table}{0}
        \renewcommand{\thetable}{S\arabic{table}}
        \setcounter{figure}{0}
        \renewcommand{\thefigure}{S\arabic{figure}}}

\newcommand{\gap}{\Delta}
\newcommand{\nep}{\mathrm{NEP}}
\newcommand{\efficiency}{\eta}
\newcommand{\rate}{\Gamma}
\newcommand{\spectralindex}{\alpha}
\newcommand{\occupancy}{n}
\newcommand{\opticalbandwidth}{B}

\newcommand{\photon}{\gamma}
\newcommand{\optical}{\mathrm{o}}
\newcommand{\recombination}{\mathrm{R}}
\newcommand{\constant}{0}
\newcommand{\tls}{\mathrm{TLS}}
\newcommand{\source}{\mathrm{S}}
\newcommand{\absorbed}{\mathrm{A}}
\newcommand{\incident}{\mathrm{I}}
\newcommand{\resonator}{\mathrm{r}}
\newcommand{\readout}{\mathrm{g}}
\newcommand{\internal}{\mathrm{i}}

\newcommand{\pairbreaking}{\mathrm{pb}}
\newcommand{\quasiparticle}{\mathrm{qp}}

\newcommand{\dark}{\mathrm{dark}}
\newcommand{\maximum}{\mathrm{max}}
\newcommand{\boltzmann}{\mathrm{B}}
\newcommand{\knee}{\mathrm{k}}
\newcommand{\cutoff}{\mathrm{c}}

\newcommand{\columbia}{Department of Physics, Columbia University, New York, NY 10027, USA}
\newcommand{\asuphysics}{Department of Physics, Arizona State University, Tempe, AZ 85287, USA}
\newcommand{\asusese}{School of Earth and Space Exploration, Arizona State University, Tempe, AZ 85287, USA}
\newcommand{\cardiff}{School of Physics and Astronomy, Cardiff University, Cardiff, Wales CF24 3AA, UK}
\newcommand{\starcryo}{STAR Cryoelectronics, Santa Fe, NM 87508, USA}
\newcommand{\jpl}{Jet Propulsion Laboratory, Pasadena, CA 91109, USA}
\newcommand{\caltech}{Division of Physics, Mathematics, and Astronomy, California Institute of Technology, Pasadena, CA 91125, USA}
\newcommand{\nrl}{Naval Research Laboratory, Washington, DC 20375, USA}

\begin{document}

\title{Photon noise from chaotic and coherent millimeter-wave sources measured with horn-coupled, aluminum lumped-element kinetic inductance detectors}

\author{D.~Flanigan}
\email{daniel.flanigan@columbia.edu}
\affiliation{\columbia}

\author{H.~McCarrick}
\affiliation{\columbia}

\author{G.~Jones}
\affiliation{\columbia}

\author{B.~R.~Johnson}
\affiliation{\columbia}

\author{M.~H.~Abitbol}
\affiliation{\columbia}

\author{P.~Ade}
\affiliation{\cardiff}

\author{D.~Araujo}
\affiliation{\columbia}

\author{K.~Bradford}
\affiliation{\asusese}

\author{R.~Cantor}
\affiliation{\starcryo}

\author{G.~Che}
\affiliation{\asuphysics}

\author{P.~Day}
\affiliation{\jpl}

\author{S.~Doyle}
\affiliation{\cardiff}

\author{C.~B.~Kjellstrand}
\affiliation{\columbia}

\author{H.~Leduc}
\affiliation{\jpl}

\author{M.~Limon}
\affiliation{\columbia}

\author{V.~Luu}
\affiliation{\columbia}

\author{P.~Mauskopf}
\affiliation{\cardiff}
\affiliation{\asusese}
\affiliation{\asuphysics}

\author{A.~Miller}
\affiliation{\columbia}

\author{T.~Mroczkowski}
\affiliation{\nrl}

\author{C.~Tucker}
\affiliation{\cardiff}

\author{J.~Zmuidzinas}
\affiliation{\jpl}
\affiliation{\caltech}

\date{\today}

\begin{abstract}
We report photon-noise limited performance of horn-coupled, aluminum lumped-element kinetic inductance detectors at millimeter wavelengths.
The detectors are illuminated by a millimeter-wave source that uses an active multiplier chain to produce radiation between 140 and 160~GHz.
We feed the multiplier with either amplified broadband noise or a continuous-wave tone from a microwave signal generator.
We demonstrate that the detector response over a 40~dB range of source power is well-described by a simple model that considers the number of quasiparticles.
The detector noise-equivalent power (NEP) is dominated by photon noise when the absorbed power is greater than approximately 1~pW, which corresponds to $\nep \approx 2 \times 10^{-17} \unit{W} \unit{Hz}^{-1/2}$, referenced to absorbed power.
At higher source power levels we observe the relationships between noise and power expected from the photon statistics of the source signal: $\nep \propto P$ for broadband (chaotic) illumination and $\nep \propto P^{1/2}$ for continuous-wave (coherent) illumination.
\end{abstract}

\maketitle

%\section{Introduction}

A kinetic inductance detector\cite{Day2003} (KID) is a thin-film superconducting resonator designed to detect photons that break Cooper pairs.
This detector technology is being developed for a range of applications across the electromagnetic spectrum.
Our devices are being developed for cosmic microwave background (CMB) studies.

The randomness of photon arrivals sets the fundamental sensitivity limit for radiation detection.
In recent years, several groups have used spectrally-filtered thermal sources to perform laboratory measurements of both aluminum and titanium nitride KIDs that demonstrate sensitivity limited by photon noise.\cite{Yates2011, Janssen2013, Mauskopf2014, deVisser2014, Hubmayr2015}
Here, we use an electronic source to demonstrate photon-noise limited performance of horn-coupled, aluminum lumped-element kinetic inductance detectors\cite{Doyle2010} (LEKIDs) sensitive to a 40~GHz spectral band centered on 150~GHz.

%\section{Experiment}

%\subsection{Detectors}

\begin{figure*}[t]
\includegraphics[width=\textwidth]{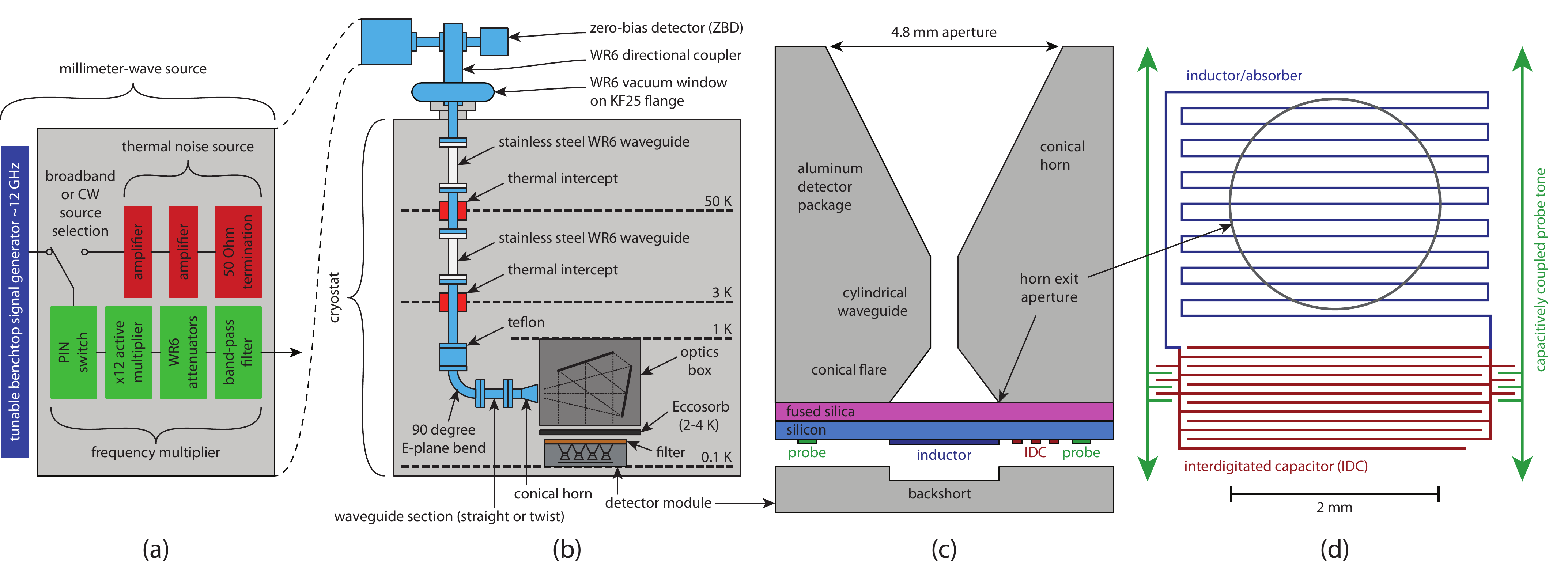}
\caption{
Experiment schematics.
\textbf{(a)} The millimeter-wave source components.
\textbf{(b)} The source and cryogenic setup.
\textbf{(c)} A cross-section of an array element. The inner conical flare and fused silica layer are designed for impedance matching.
\textbf{(d)} The lumped circuit elements of one LEKID.
Parts of this figure are reproduced with permission from H.~McCarrick \textit{et al.}, Rev. Sci. Instrum. \textbf{85}, 123117 \copyright 2014 American Institute of Physics.
}
\label{fig:experiment}
\end{figure*}

The array of devices used in this study was fabricated by patterning a 20~nm aluminum film on a high-resistivity crystalline silicon substrate, with twenty detectors per array.
Each resonator comprises lithographed structures that behave electrically as lumped elements, namely an interdigitated capacitor and an inductive meander that is also the photon absorber.
Schematics of a detector and the horn coupling scheme are shown in Figure~\ref{fig:experiment}.
These devices were fabricated at STAR Cryoelectronics using the same lithographic mask used to pattern the devices described in a previous study.\cite{McCarrick2014}
The same processing steps were used in this study except that the silicon wafer was immersed in hydrofluoric acid prior to aluminum deposition in order to clean and hydrogen-terminate the silicon surface to reduce oxide formation.
We measure a superconducting transition temperature $T_c = 1.39 \unit{K}$.
The resonance frequencies are $95 \unit{MHz} < f_\resonator < 195 \unit{MHz}$.
Under the lowest loading conditions the internal quality factors are $Q_\internal \approx 5 \times 10^5$.
The coupling quality factors are $Q_c \approx 5 \times 10^4$.
The volume of each inductive meander is $1870 \unit{\mu m}^3$, assuming nominal film thickness.
The detector bath temperature is $120 \pm 1 \unit{mK}$, obtained in a cryostat using an adiabatic demagnetization refrigerator backed by a helium pulse tube cooler.
Detector readout is performed with a homodyne system using a cryogenic SiGe low-noise amplifier and open-source digital signal-processing hardware.\cite{McCarrick2014, ColumbiaCMB}
All the data shown are from a single representative detector with $f_\resonator = 164 \unit{MHz}$, and were taken at a constant readout tone power of approximately $-100 \unit{dBm}$ on the feedline.
The package that contains the detector chip is machined from QC-10, which is an aluminum alloy known to superconduct at the bath temperature used here.

%\subsection{Millimeter-wave source}

Figure~\ref{fig:experiment}(a) is a schematic of the millimeter-wave source, located outside the cryostat.
Within the source, the output of a $12\times$ active multiplier chain passes through two variable waveguide attenuators that allow the output power to be controlled over a range of more than 50 dB.
The primary components of the source are listed in the Supplemental Material.\cite{supplement}

The output spectrum is controlled by a band-pass filter with a sharp roll-off outside its passband of 140 to 160~GHz.
Within this passband, the source can produce radiation in two modes.
In \textit{broadband} mode, amplified noise is multiplied into a broadband chaotic signal.
In \textit{continuous-wave} mode, a multiplied tone from a signal generator approximates a monochromatic coherent signal.
We have measured the source output in both modes using a Fourier transform spectrometer; these measurements show that in broadband mode the power is constant within a factor of two across the output band, and in continuous-wave mode it appears monochromatic with negligible higher harmonics.

Figure~\ref{fig:experiment}(b) shows the signal path from the source through the cryostat to the detectors.
The source output is split using a waveguide directional coupler that sends 99\% of the power into a calibrated, isolator-coupled zero-bias diode power detector (ZBD), the voltage output of which is recorded using a lock-in amplifier.
The remaining 1\% of the power travels through a vacuum window and into the cryostat through WR6 waveguide.
A piece of Teflon at 4~K inserted into the waveguide absorbs room-temperature thermal radiation.
Two mirrors transform the output of a conical horn into a collimated beam.
A 6.4~mm thick slab of microwave absorber (Eccosorb MF-110), regulated at 2~K during these measurements, attenuates incoming signals and provides a stable background load.
A metal-mesh filter at the detector apertures defines the upper edge of the detector band at 170~GHz.
The lower edge of the band at 130~GHz is defined by the cutoff frequency of a 1.35~mm diameter circular waveguide in the detector package.
We note that the source output is within the single-mode bandwidth of both WR6 waveguide and the circular waveguide.
The radiation from the source incident on the detector horns is linearly polarized, and the electric field is aligned with the long elements of the inductive meanders in the detectors.

%\section{Results}

\begin{figure*}[t]
\includegraphics[width=0.9\textwidth]{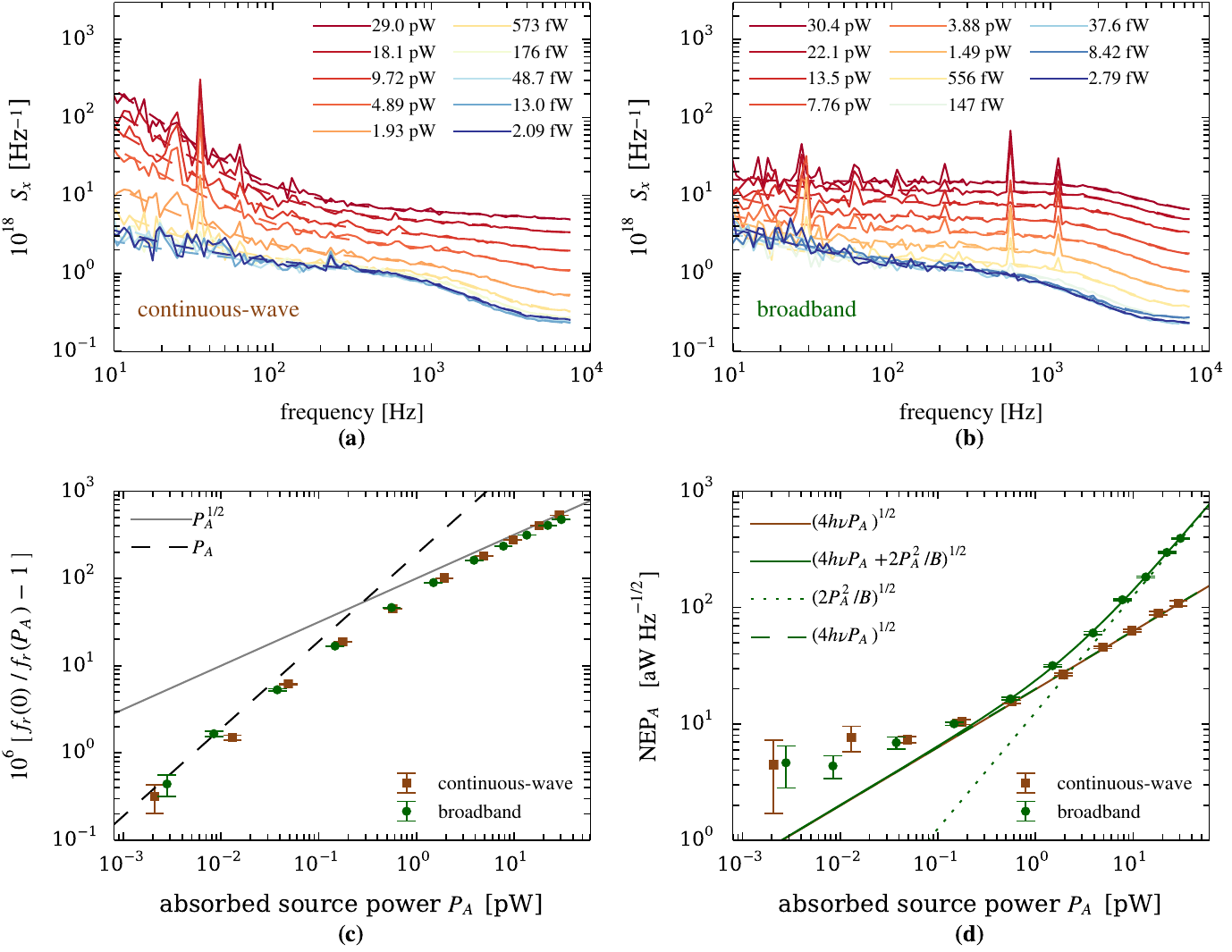}
\caption{
Primary results of the experiment.
\textbf{(a)} Spectral density $S_x$ of detector time-ordered data versus frequency under continuous-wave illumination with $\nu = \fcw \unit{GHz}$ (solid lines), and the result of fitting the data to Equation~\ref{eqn:noise_model} (dashed lines).
At high power the red noise component is dominated by fluctuations from the signal generator that feeds the multiplier; these fluctuations are correlated among detectors.
\textbf{(b)} Spectral density under broadband illumination, and fits of Equation~\ref{eqn:noise_model}.
The spikes above 400~Hz are pickup from a fan in the source.
The red noise below $100 \unit{Hz}$ at low source power in both modes is produced by vibrations from the pulse tube cooler that vanish when it is turned off.
The detector white noise levels from the fits are used to calculate NEP values.
\textbf{(c)} Fractional frequency response versus absorbed power in both source modes.
The error bars are statistical errors from the resonator fits.
We use the finite-difference derivative of these response data to calculate the NEP.
The dashed black line and solid gray line are guides that show how the response scales at both low and high absorbed power.
\textbf{(d)}
Noise-equivalent power versus absorbed power in both source modes.
All data points and lines are referenced to absorbed power.
The error bars are propagated statistical errors from the finite difference derivative and the detector noise fits.
The solid green line is the sum of the quadratic and linear terms in the fit of Equation~\ref{eqn:nep} to the broadband $\nep^2$ data.
The dotted green line is the quadratic term, which is the photon wave noise contribution.
The dashed green line is the linear term, which contains equal contributions from photon shot noise and quasiparticle recombination noise.
The broadband frequency used is $\nu = 150 \unit{GHz}$, near the band center.
The solid brown line (nearly coincident with dashed green) is the linear term in the fit of Equation~\ref{eqn:nep} to the continuous-wave $\nep^2$ data, in which the quadratic term is omitted.
}
\label{fig:results}
\end{figure*}

Figure~\ref{fig:results} shows the main results of this work.
All power values in this figure refer to the power from the source absorbed by the detector: $P_\absorbed = \efficiency_\source P_\source$, where $P_\source$ is measured by the ZBD.
Before calibration, the efficiency $\efficiency_\source$ is known only approximately from measurements and simulations of the components between the source and the detector.
We accurately determine $\efficiency_\source$, and thus the absorbed source power, by measuring the relationship between emitted source power and detector noise.
This calibration relies on the assumption that all components between the source output and detector are linear: we have linearized the ZBD response at the higher power levels, all other components are passive, and we assume that filter heating is negligible.
To perform the calibration we use measurements of the noise-equivalent power (NEP), defined as the standard error of the mean in the inferred optical power at a given point in the optical system after 0.5~s of integration.\cite{Richards1994, Zmuidzinas2003}
We calculate the NEP using measurements of the detector noise and responsivity.

%\subsection*{Detector response}

At each source power level, to determine the resonance frequency and the quality factors we sweep the readout tone generator frequency $f_\readout$ across a resonance and fit a resonator model to the forward scattering parameter $S_{21}(f_\readout)$ data.\cite{McCarrick2014}
Figure~\ref{fig:results}(c) shows the detector response to source power in both broadband and continuous-wave modes.
At low source power in both modes the fractional frequency shift $x(P_\absorbed) = f_\resonator(0) / f_\resonator(P_\absorbed) - 1$ is approximately linear in power, while at high power $x \propto P_\absorbed^{1/2}$.
This behavior is described by a model in which the fractional frequency shift is proportional to the number of quasiparticles:
\begin{equation}
\label{eqn:quasiparticle_number}
N_\quasiparticle
  =
  N_* \left[
  \left( 1 + 2 \tau_{\mathrm{max}} (\rate_\constant + \rate_\source) / N_* \right)^{1/2} -1
  \right].
\end{equation}
Here, $\rate_\source \propto P_\absorbed$ is the rate of quasiparticle generation due to absorbed source photons, $\rate_\constant$ is the constant generation rate due to other effects (such as absorption of ambient photons and thermal phonons), and $N_*$ and $\tau_{\mathrm{max}}$ are material-dependent constants that describe the observed saturation of the quasiparticle relaxation time at low quasiparticle number,\cite{Zmuidzinas2012} with
$\tau_\quasiparticle \approx \tau_\mathrm{max} / (1 + N_\quasiparticle / N_*)$.
(This saturation is not experimentally accessible here.)
We calculate the responsivity $\mathrm{d} x / \mathrm{d} P_\source$ at each source power level with a finite-difference derivative that uses the fractional frequency response at adjacent power levels.

%\subsection*{Detector noise}

To measure detector noise we record time-ordered data $S_{21}(f_\readout = f_\resonator)$.
Using the resonator model from the fit to the frequency sweep we convert these data into units of fractional frequency shift $x$, then calculate the single-sided spectral density $S_x(f)$.
Figures~\ref{fig:results}(a) and \ref{fig:results}(b) show the measured noise spectra and fits to the following model:\cite{supplement}
\begin{equation}
\label{eqn:noise_model}
S_x(f)
  =
  W^2 \frac{1 + (f_\knee / f)^\spectralindex}{1 + (f / f_\cutoff)^2} + A^2,
\end{equation}
where the free parameters are the detector white noise $W^2$, the red noise knee frequency $f_\knee$, the spectral index $\spectralindex$, the cutoff frequency $f_\cutoff$, and the amplifier noise $A^2$.
This model treats the detector noise as the sum of a white noise process with spectral density $W^2$ and a red noise process with spectral density $R^2 = W^2 (f_k / f)^\spectralindex$, both rolled off at $f_\cutoff$.

The detector audio bandwidth of about $1 \unit{kHz}$ corresponds to a limiting time constant $\tau = (2 \pi f_\cutoff)^{-1}$ that is approximately equal to both the resonator ring-down time $\tau_\resonator = Q / \pi f_\resonator$ and the expected quasiparticle relaxation time $\tau_\quasiparticle$ for aluminum.
Both of these time constants are expected to decrease as the absorbed optical power increases, as observed in the data.

To model the detector noise, we first consider noise sources independent of the quasiparticle system.
White noise due to the cryogenic amplifier dominates at frequencies well above the detector bandwidth, and we account for it in the noise spectra model.
Two-level systems (TLS) in amorphous dielectric surface layers located near the resonator produce fluctuations in the local dielectric constant and thus in $f_\resonator$.\cite{Gao2008}
In a separate experiment, described in the Supplemental Material,\cite{supplement} we determined that TLS noise is negligible at the readout power level (-100~dBm) used in the measurements presented here and thus do not include it in the noise model.
The chosen readout power level is high enough to suppress TLS noise but is not so high that nonlinear effects due to resonator bifurcation become significant.

The remaining noise sources involve fluctuations in the quasiparticle system: generation by optical photons, readout photons, and thermal phonons, as well as quasiparticle recombination, e.g. via phonon emission.
All of these sources are expected to produce white noise that rolls off at the frequency corresponding to the larger of $\tau_\resonator$ and $\tau_\quasiparticle$.\cite{Zmuidzinas2012}
We expect readout generation to be negligible at high source power, and treat it as constant.
(Where present, the photon wave noise introduces correlations between photon arrival times.
This noise has a bandwidth equal to the 20~GHz bandwidth of the absorbed broadband radiation, so it is also expected to appear white in the detector audio band.\cite{Zmuidzinas2015})

% \subsection*{NEP model}

The NEP model includes theoretical expectations for photon noise and quasiparticle recombination noise.
We denote by $\occupancy$ the mean photon occupancy of a single spatial/polarization mode of the electromagnetic field with frequency $\nu$.
For example, for a thermal source at temperature $T$ the occupancy is
$\occupancy = [\exp(h \nu / k_\boltzmann T) - 1]^{-1}$,
where $h$ is Planck's constant and $k_\boltzmann$ is Boltzmann's constant.
If we assume that the radiation occupies an effective optical bandwidth $\opticalbandwidth \ll \nu$ sufficiently narrow that quantities such as occupancy and absorption efficiency can be treated as constant, then the power from this mode that is absorbed by a detector with absorption efficiency $\efficiency$ is
$P_\absorbed = \efficiency \occupancy \opticalbandwidth h \nu$.
If the source is thermal then the contribution of photon noise to the NEP is given by\cite{Zmuidzinas2003}
\begin{equation}
\label{eqn:photon_nep}
\nep_{\absorbed, \photon}^2
  =
  2 \efficiency \occupancy (1 + \efficiency \occupancy) \opticalbandwidth (h \nu)^2
  =
  2 h \nu P_\absorbed + 2 P_\absorbed^2 / \opticalbandwidth,
\end{equation}
which is referenced to absorbed power.
We refer respectively to these two terms as shot noise and wave noise, following Hanbury Brown and Twiss.\cite{HBT1957}
If the source is monochromatic with perfect temporal coherence then only the shot noise term is present regardless of the occupancy: this behavior represents a key difference between a quantum coherent state and a quantum-statistical thermal state of the field.\cite{Loudon2002, Glauber2005}
For a thermal source, if $\efficiency \occupancy \ll 1$ the shot noise dominates, which is typical in optical astronomy; if $\efficiency \occupancy \gg 1$ the wave noise dominates, which is typical in radio astronomy.

We measure power at the output of the source and detector NEP referenced to the same point.
Referencing the photon $\nep$ to the source output gives
\begin{equation}
\label{eqn:photon_nep_source}
\nep_{\source, \photon}^2
  =
  \nep_{\absorbed, \photon}^2 / \efficiency_\source^2
  =
  2 h \nu P_\source / \efficiency_\source + 2 P_\source^2 / B.
\end{equation}
The presence of the efficiency $\efficiency_\source$ in the linear term of this equation enables extraction of the absorbed source power.

Previous studies that calculated the absorption efficiency of a KID by measuring the scaling of photon shot noise with optical power have used superconducting films with transition temperatures similar to the film used here but larger photon energies.\cite{Yates2011, Janssen2013, deVisser2014, Hubmayr2015}
Here, the photons have energies $h \nu \gtrsim 2 \gap$, where $\gap$ is the superconducting energy gap, so each photon excites only two quasiparticles close to the gap; in this limit the quasiparticle recombination noise is significant.
The recombination noise contribution to $\nep_\absorbed$ is\cite{supplement}
\begin{equation}
\label{eqn:recombination_nep}
\nep_{\absorbed, \recombination}^2
  =
  4 \Delta P_\absorbed / \efficiency_\pairbreaking
\end{equation}
where $\efficiency_\pairbreaking$ is the pair-breaking efficiency.
For photon energies $2 \gap < h \nu < 4 \gap$, a recent measurement\cite{deVisser2015} found $\efficiency_\pairbreaking \approx 2 \gap / h \nu$, in agreement with theory.\cite{Guruswamy2014}
Using this value, the recombination NEP equals the shot noise term in the photon NEP.
This is expected based on the symmetry between uncorrelated pair-breaking events and uncorrelated pair-recombination events.
Finally, we introduce a small constant term $\nep_\constant$ to account for noise sources independent of source power, such as TLS noise and quasiparticle generation-recombination noise from thermal phonons, readout photons, and ambient photons.

To calculate the detector $\nep_\absorbed$, which is shown in Figure~\ref{fig:results}(d), we use the measured fractional frequency shift $x$ (unitless), the measured fractional frequency noise power $S_x$ (1 / Hz), and the source power $P_\source$ (watts) as measured with a calibrated zero-bias diode (ZBD) mounted on the directional coupler outside the cryostat (see Figure 1).
The source power absorbed by the detector is related to $P_\source$ by $P_\absorbed = \efficiency_\source P_\source$ where $\efficiency_\source$ is an overall system efficiency from the source output to the detector that includes the transmission through the directional coupler, the attenuation of the stainless steel waveguide, the geometrical dilution due to the internal optics, the loss in the Eccosorb, and the detector absorption efficiency.
To compute the responsivity to changes in the source power, we plot $x$ versus $P_\source$ and calculate the slope of this curve $\mathrm{d} x / \mathrm{d} P_\source$ at each $P_\source$ using a finite difference algorithm.
We use this responsivity to convert the fractional frequency noise measurements ($S_x$) to $\nep_\source$.
Note that for $\nep_\source$ we use only the white noise component, $W$, obtained by fitting Equation~\ref{eqn:noise_model} to each $S_x$ measurement.
Thus, $\nep_\source = W / (\mathrm{d} x / \mathrm{d} P_\source)$.
To convert $P_\source$ to $P_\absorbed$ we need to determine $\efficiency_\source$.
The complete theoretical model for $\nep_\source$ is
\begin{align}
\begin{split}
\label{eqn:nep}
\nep_\source^2
  &=
  (\nep_{\absorbed, \constant}^2
   + \nep_{\absorbed, \recombination}^2
   + \nep_{\absorbed, \photon}^2) / \efficiency_\source^2 \\
  &=
  \nep_{\absorbed, \constant}^2 / \efficiency_\source^2
  + [2 (2 h \nu P_\absorbed) + 2 P_\absorbed^2 / B] / \efficiency_\source^2 \\
  &=
  \nep_{\source, \constant}^2
  + 4 h \nu P_\source / \efficiency_\source
  + 2 P_\source^2 / B,
\end{split}
\end{align}
which is the sum of the aforementioned noise contributions.
The right-hand side of this equation is quadratic in $P_\source$ with unknown quantities $\nep_{\source, \constant}$, $\efficiency_\source$, and effective optical bandwidth $B$.
The limiting $\nep_{\source, \constant}$ is discussed below.
We fit Equation~\ref{eqn:nep} to the broadband data using center frequency $\nu = 150 \unit{GHz}$ and obtain $\efficiency_\source = 8.50 \times 10^{-7} (1 \pm 0.09)$ and $\opticalbandwidth = 13 \unit{GHz}$.
The quadratic term is not expected to be present for coherent
illumination because the source should produce only shot noise, so we fit Equation~\ref{eqn:nep} to the continuous-wave data omitting the third term.
Here, $\nu = \fcw \unit{GHz}$ and we obtain $\efficiency_\source = 1.12 \times 10^{-6} (1 \pm
0.04)$.
As a final step, we convert $P_\source$ to $P_\absorbed$ using the $\efficiency_\source$ values from the model fitting and produce Figures~\ref{fig:results}(c) and \ref{fig:results}(d).
Note that because the broadband source involves contributions from the full source output bandwidth, it is not surprising that the measured $\efficiency_\source$ values differ between the continuous-wave and broadband modes by more than the statistical error bars.

%\section{Discussion}

Figure~\ref{fig:results}(d) shows that photon noise dominates under broadband illumination when
$P_\absorbed \gtrsim 1 \unit{pW}$,
which corresponds to 
$\nep_\absorbed \approx 2 \times 10^{-17} \unit{W} \unit{Hz}^{-1/2}$.
At high power in each source mode we observe the expected relationship between noise and power: in broadband mode $\nep \propto P$ because the quadratic wave noise term dominates, while in continuous-wave mode $\nep \propto P^{1/2}$ because the quadratic term is not present.
This behavior is a clear signature of photon noise.

Note that the the $\nep_\absorbed$ values reported have the amplifier noise contribution subtracted because the white noise parameter $W^2$ in Equation~\ref{eqn:noise_model} describes the noise power above the amplifier noise $A^2$.
Here, subtracting the amplifier noise yields an accurate estimate of the detector performance because, alternatively, the amplifier noise can be suppressed to a negligible level by increasing the readout power.
We verified both approaches yield the same $\nep_\absorbed$ versus $P_\absorbed$ result but chose to report the amplifier-noise-subtracted results.

At low absorbed source power levels in both modes, where $P_\absorbed < 0.1 \unit{pW}$, $\nep_\absorbed$ levels off to $\nep_\constant$.
The values of $\nep_\constant$ extracted from both of the aforementioned fits are approximately 5 to $6 \times 10^{-18} \unit{W} \unit{Hz}^{-1/2}$.
To explain this leveling-off effect, we model the background loading as emission from a black body at 2~K, which is the temperature of the Eccosorb in front of the feed horn apertures.
Assuming center frequency $\nu = 150 \unit{GHz}$, measured filter transmission $\efficiency_\mathrm{F}(\nu) = 0.94$, optical efficiency $\efficiency_\incident = 0.7$ (obtained from electromagnetic simulations), and detector bandwidth $\opticalbandwidth_\mathrm{full} = 40 \unit{GHz}$, then the radiative loading from the Eccosorb is $P_\absorbed = \efficiency_\incident \occupancy(\nu, 2 \unit{K}) h \nu \opticalbandwidth_\mathrm{full} = 0.08 \unit{pW}$.
This loading level is close to the observed knee in the curves in Figure~\ref{fig:results}(d).
Adding an equal recombination noise contribution to the corresponding photon NEP gives $\nep_\absorbed = (2 \cdot 2 h \nu P_\absorbed)^{1/2} = 5.6 \times 10^{-18} \unit{W} \unit{Hz}^{-1/2}$, which is close to the observed $\nep_\constant$ value.
Therefore, the observed limiting $\nep_\absorbed$ is consistent with this expected background loading model.

Analysis of data from twelve detectors yielded similar results to
those shown in Figure~\ref{fig:results}(d), with the photon noise starting to dominate between 0.5 and 1~pW.
We conclude that these detectors become limited by photon noise at absorbed power levels lower than the background power levels already measured by ground-based CMB polarimeters.

%\section*{Acknowledgements}
\bigskip

R. C. is both an author and the owner of STAR Cryoelectronics.
H. M. is supported by a NASA Earth and Space Sciences Fellowship.
T. M. is supported by a National Research Council Fellowship.
This research is supported in part by a grant from the Research Initiatives for Science and Engineering program at Columbia University to B. R. J., and by internal Columbia University funding to A. M.
We thank the Xilinx University Program for their donation of FPGA hardware and software tools used in the readout system.
We thank the anonymous reviewers for thoughtful and helpful comments.

\bibliography{references.bib}
\clearpage

\beginsupplement
\section*{Supplemental Material}

\subsection*{Millimeter-wave source}

\begin{table}[!h]
\centering
\caption{Primary components of the millimeter-wave source.}
\renewcommand{\arraystretch}{1.2}
\begin{tabular}{lll}
\toprule
\textbf{Component} & \textbf{Vendor} & \textbf{Part Number} \\
\midrule
50~$\Omega$ terminator & Minicircuits & ANNE-50X \\ 
High gain amplifiers & Spacek Labs & SG134-40-17 \\ 
PIN switch & Narda & S213D \\ 
Active multiplier & Millitech & AMC-05 \\
Variable attenuators & Custom Microwave & VA6R \\
Band-pass filter & Pacific Millimeter & 14020 \\
Directional coupler & Millitech & CL3-006 \\
Zero-bias diode power detector & Virginia Diodes, Inc. & WR6.5-ZBD \\
\bottomrule
\end{tabular}
\end{table}

\subsection*{Two-level system noise}

At low temperatures we see evidence for two-level system (TLS) effects in measurements of resonance frequency versus bath temperature, which depart from the Mattis-Bardeen prediction, and in the fact that the internal quality factors increase with increasing readout power.
The connection between these static TLS effects and TLS noise is not fully understood.
The method we used to estimate the TLS noise contribution is described in this section.
We conclude that TLS noise is negligible and thus do not include it explicitly in the analysis of the NEP.

In this work, the motivation for modeling TLS noise is that the detector responsivity decreases with increased optical loading: as shown in Figure~2(c), at high power $x(P_\source) \propto P_\source^{1/2}$, so the responsivity  $\mathrm{d} x / \mathrm{d} P_\source \propto P_\source^{-1/2}$.
Thus, a noise source with constant amplitude $S$ in fractional frequency units would be linear in power when converted to NEP units:
\begin{equation*}
\nep^2
  =
  S (\mathrm{d} x / \mathrm{d} P_\source)^{-2} \propto S P
\end{equation*}
The presence of such a noise source would complicate the extraction of the linear NEP term.

The TLS contribution to the spectral density is typically
\begin{equation*}
S_{x, \tls}(f, P_\internal)
  \propto
  f^{-1/2} (1 + P_\internal / P_*)^{-1/2},
\end{equation*}
where $P_\internal$ is the internal readout power and the critical power $P_*$ is small compared to the readout power levels typically used with KIDs. \cite{Gao2007, Gao2008, Barends2010, Zmuidzinas2012, Neill2013}
The experiment described in the main text is performed with constant readout power $P_\readout$ on the feedline, and we expect the TLS noise level to vary as $P_\internal^{-1/2} = (\chi_a P_\readout)^{-1/2}$, where $\chi_a \le 1/2$ can be calculated from resonator parameters.\cite{Zmuidzinas2012}

\begin{figure}[t!]
\includegraphics[width=\columnwidth]{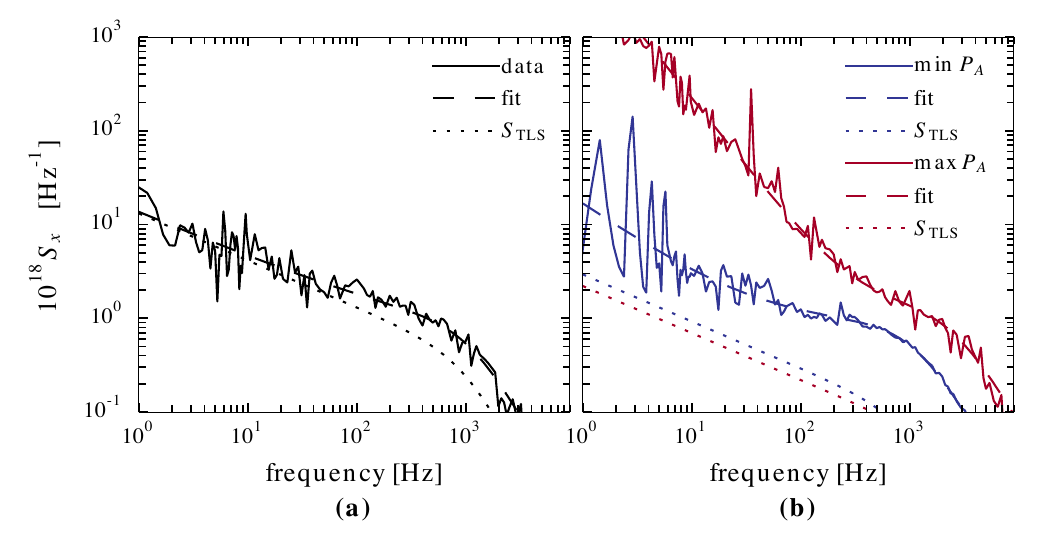}
\caption{
\textbf{(a)}
Amplifier-subtracted dark noise data for the same detector characterized in the main text.
The dashed line shows a fit to the same model used in the main text, except that here the spectral index is fixed to $\alpha = 0.5$ to match a possible TLS contribution.
To show the detector noise more clearly, the amplifier noise value obtained from the fit has been subtracted from the data and fit curves.
The dotted line shows the possible TLS contribution, assumed to roll off with the same time constant obtained from the fit.
\textbf{(b)}
Amplifier-subtracted illuminated continuous-wave noise data.
The solid lines shown here are the lowest and highest power curves from Figure~2(a), and the dashed lines are the same fits shown in the main text, except that the amplifier noise values obtained from the fits have been subtracted from the data and fit curves.
The dotted lines are the inferred TLS contribution to the illuminated spectra, scaled from the fit value in panel (a) by a factor $(P_{\internal, \dark} / P_\internal)^{1/2}$.
The TLS contribution in this case decreases as source power increases.}
\label{fig:dark_noise}
\end{figure}

In order to estimate the TLS contribution to the NEP, we performed a separate experiment in which we attempted to make any TLS noise as prominent as possible.
Three key aspects differ from the experiment described in the main text: the horn apertures were covered with aluminum tape to minimize optical loading; the readout power was approximately $-112 \unit{dBm}$, 12~dB lower than in the primary experiment; in order to remove noise due to vibrations caused by the pulse tube cooler, we turned it off to record time-ordered data while the adiabatic demagnetization refrigerator continued to regulate the bath temperature at 120~mK.

Figure~\ref{fig:dark_noise}(a) shows the fractional frequency spectral density taken under these dark conditions and a fit used to extract a possible TLS noise contribution.
Figure~\ref{fig:dark_noise}(b) shows that this TLS contribution is negligible when adjusted for the increased readout power used in the primary experiment.

\subsection*{Recombination noise}

For clarity, in this section we include a derivation of Equation~5 from the main text because it differs by a factor of two from expressions that have appeared in the MKID literature.\cite{Yates2011, Janssen2013, deVisser2014, McCarrick2014, Hubmayr2015}

Consider a flow of quanta $q$ at rate $\rate_q$ events per second.
Assume that the current $I_q = q \rate_q$ is stationary and that the events are uncorrelated.
Then, for positive frequencies the single-sided spectral density of the current is constant:
\begin{equation*}
S_{I_q} = 2 q I_q = 2 q^2 \rate_q,
\end{equation*}
with units of current squared per hertz.
(All the spectral densities written here are single-sided.)
If photons from a coherent source with frequency $\nu > 2 \gap / h$ are absorbed in a detector at rate $\rate_\nu$ then the photon arrival times are uncorrelated, the quantum is one photon, and the spectral density of the absorbed photon flux is
$S_{\rate_\nu} = 2 \rate_\nu$.

Let $m \ge 2$ be the mean number of quasiparticles produced per absorbed photon so that $\rate_\optical = m \rate_\nu$ is the quasiparticle generation rate due to optical photons.
The relationship between $m$ and the pair-breaking efficiency $\efficiency_\pairbreaking$ is
\begin{equation*}
m
  =
  \frac{\efficiency_\pairbreaking P_\absorbed}{\rate_\nu \gap}
  =
  \frac{\efficiency_\pairbreaking h \nu}{\gap},
\end{equation*}
since $P_\absorbed / \gap$ would be the generation rate if all of the power excited quasiparticles exactly at the gap.
Again considering a coherent source, the spectral density of the generation rate due to absorbed photons is
\begin{equation*}
S_{\rate_\optical}
  =
  2 m^2 \rate_\nu
  =
  2 m^2 (\rate_\optical / m)
  =
  2 m \rate_\optical.
\end{equation*}
Two quasiparticles recombine per event, so the spectral density of the recombination rate $\rate_\recombination$ is
\begin{equation*}
S_{\rate_\recombination}
  =
  2 \; (2)^2 \; (\rate_\recombination / 2)
  =
  4 \rate_\recombination.
\end{equation*}
The rate equation for the quasiparticle number is
$\mathrm{d} N_\quasiparticle / \mathrm{d} t = \rate - \rate_\recombination$, where $\rate$ is the total rate at which quasiparticles are excited and $\rate_\recombination$ is the rate at which they recombine.
In steady state, the recombination rate due to optically excited photons equals the optical generation rate.
(If optically excited quasiparticles dominate then we can also neglect recombination due to other excitation sources.)
We see that if $\rate_\optical = \rate_\recombination$
then
$S_{\rate_\optical} / S_{\rate_\recombination} = m / 2$.
In this work $m = 2$, so the generation noise from a coherent optical source equals the recombination noise due to the quasiparticles it excites.

To connect the above spectral densities to NEP referenced to a given point in an optical system, relate $\rate_\optical$ to incident power $P$ at that point:
\begin{equation*}
P
  =
  \frac{P_\absorbed}{\efficiency}
  =
  \frac{h \nu \rate_\nu}{\efficiency}
  =
  \frac{h \nu \rate_\optical}{m \efficiency},
\end{equation*}
where $P_\absorbed = \efficiency P$.
Referencing $S_{\rate_\optical}$ to incident power gives
\begin{equation*}
\nep^2
  =
  \left( \frac{h \nu}{m \efficiency} \right)^2 S_{\rate_\optical}
  =
  \frac{2 h \nu P}{\efficiency},
\end{equation*}
which matches the shot noise term in Equation~4 if the reference point is the source output.
Referencing the recombination noise to incident power in the same way gives
\begin{equation*}
\nep_\recombination^2
  =
  \left( \frac{h \nu}{m \efficiency} \right)^2 S_{\rate_\recombination}
  =
  \frac{2}{m} \frac{2 h \nu P}{\efficiency}.
\end{equation*}
Using $\efficiency_\pairbreaking$ instead of $m$ gives
\begin{equation*}
\nep_\recombination^2
  =
  \frac{4 \gap P}{\efficiency_\pairbreaking \efficiency},
\end{equation*}
which matches Equation~5 if multiplied by $\efficiency^2$ to reference to absorbed power.
(This equation for recombination NEP appears in Equations~95 and 97 of \citet{Zmuidzinas2012} with $\efficiency_\pairbreaking \rightarrow \efficiency_o$ and $\efficiency = 1$.)

The quasiparticle recombination rate is\cite{Zmuidzinas2012}
\begin{equation*}
\rate_\recombination
  =
  \frac{N_\quasiparticle}{2} \left( \tau_\quasiparticle^{-1} + \tau_\maximum^{-1} \right)
  =
  \frac{N_\quasiparticle^2}{2 N_* \tau_\maximum} + \frac{N_\quasiparticle}{\tau_\maximum},
\end{equation*}
with relaxation time $\tau_\quasiparticle \le \tau_\maximum$.
For simplicity, assume that the only relevant quasiparticle decay process is recombination with phonon emission so that $\tau_\quasiparticle \ll \tau_\maximum$ and we can neglect the second terms on both sides of the second equality above. 
Then, the low-frequency  ($2 \pi f \tau_\quasiparticle \ll 1$) contribution of recombination to the spectral density of the quasiparticle number is
\begin{equation*}
S_{N_\quasiparticle, \recombination}
  =
  S_{\rate_\recombination} \left( \mathrm{d} \rate_\recombination / \mathrm{d} N_\quasiparticle \right)^{-2}
  =
  4 \rate_\recombination \left( 2 \rate_\recombination / N_\quasiparticle \right)^{-2}
  =
  2 N_\quasiparticle \tau_\quasiparticle.
\end{equation*}
If the generation noise equals the recombination noise, as in thermal equilibrium without optical generation, then multiplying the above by two gives $S_{N_\quasiparticle} = 4 N_\quasiparticle \tau_\quasiparticle$.
This is consistent with Wilson and Prober,\cite{Wilson2004} who derive the single-sided spectral density of the quasiparticle number
$G(\omega)
  =
  4 N_0 \tau_r^* (1 + (\omega \tau_r^*)^2)^{-1}$,
where their $N_0 = N_\quasiparticle$ and $\tau_r^*$ is the effective relaxation time for perturbations, as is $\tau_\quasiparticle$ here.

Using the recombination rate to write the recombination NEP in terms of the quasiparticle number gives
\begin{equation*}
\nep_\recombination^2
  =
  \left( \frac{\gap}{\efficiency_\pairbreaking \efficiency} \right)^2
  4 \frac{N_\quasiparticle}{2 \tau_\quasiparticle}
  =
  \frac{2 \gap^2 N_\quasiparticle}{\efficiency_\pairbreaking^2 \efficiency^2 \tau_\quasiparticle}
\end{equation*}
which also matches earlier work.\cite{Yates2011, deVisser2014, Zmuidzinas2012}
(In the $\tau_\quasiparticle \ll \tau_\maximum$ regime, the equation for the power flow through the steady-state quasiparticle system is $\efficiency_\pairbreaking \efficiency P = \efficiency_\pairbreaking P_\absorbed = \gap \rate_\recombination = \gap N_\quasiparticle / 2 \tau_\quasiparticle$, as required for consistency between these equations for $\nep_\recombination^2$.)
We conclude that only Equation~5 in the main text is discrepant with some of the previous MKID literature.
Because we use this equation as part of the NEP model, we are unable to empirically demonstrate that the equation given here is correct.

Finally, the photon NEP of a chaotic source will also include a wave noise term. This does not affect the conclusion that, in this work, the photon shot NEP term equals the NEP due to recombination of quasiparticles excited by that source.

\subsection*{Spectral density fitting}

In this section we provide details of the procedure used to fit the spectral density to Equation~2 of the main text.
To estimate the spectral density of the time-ordered fractional frequency shift data we first use Welch's average periodogram method with the data split into 16 equal non-overlapping chunks.
This produces a single-sided spectral density that is the average of 16 spectra.
We estimate the variance of point $j$ with value $S_j$ by $\sigma_j^2 = S_j^2 / 16$.
We then bin this spectrum using bin widths that increase with frequency, and propagate the errors by adding the variances in quadrature.
These binned spectra are plotted in Figures~2(a) and 2(b).

This binning and averaging procedure produces $\chi^2$ distributed data with $2 \times 16 \times n_k$ degrees of freedom, where $n_k$ is the number of points that are averaged in bin $k$.
The resulting distribution closely approximates a Gaussian distribution,\cite{Norrelykke2010} even for $n_k = 1$.
To fit the model to the data we use a least-squares fitting routine with the squared residual at each frequency point weighted by the inverse of the variance in that bin.
Only data at frequencies above 10~Hz is used in the fits.
This model will over-describe the data if the spectrum has no red noise or no white noise component, in which case the uncertainties on the remaining parameters are underestimated.
The resulting best-fit parameters are listed in Tables~\ref{tab:bbfitparams}~and~\ref{tab:cwfitparams}.

\begin{table}[tbp]
\centering
\caption
[Best-fit parameters from the spectral density fits of the broadband data.]
{Broadband best-fit parameters with uncertainties.
At high power, because the noise is very close to white, the red noise contribution is negligible and the parameters ($f_\knee$ and $\spectralindex$) that describe the red noise are poorly constrained; the white noise $W^2$ is still well constrained.}
\renewcommand{\arraystretch}{1.2}
\begin{tabular}{cccccc}
\toprule
$P_A$ & $A^2 \; [\SI{e-18}{Hz^{-1}}]$ & $W^2 \; [\SI{e-18}{Hz^{-1}}]$ & $f_\knee \; [\si{Hz}]$ & $\spectralindex$ & $f_\cutoff \; [\si{kHz}]$ \\
\midrule
\SI{30.4}{pW} & 5.2 $\pm$ 0.1 & 10.0 $\pm$ 0.2 & 0 $\pm$ 20 & 1 $\pm$ 6 & 3.0 $\pm$ 0.1 \\
\SI{22.1}{pW} & 3.9 $\pm$ 0.1 & 7.7 $\pm$ 0.1 & 11 $\pm$ 3 & 1.9 $\pm$ 0.9 & 2.9 $\pm$ 0.1 \\
\SI{13.5}{pW} & 2.67 $\pm$ 0.06 & 5.3 $\pm$ 0.1 & 7 $\pm$ 4 & 1.0 $\pm$ 0.5 & 2.7 $\pm$ 0.1 \\
\SI{7.76}{pW} & 1.38 $\pm$ 0.03 & 3.81 $\pm$ 0.08 & 12 $\pm$ 3 & 1.2 $\pm$ 0.4 & 2.48 $\pm$ 0.08 \\
\SI{3.88}{pW} & 0.82 $\pm$ 0.02 & 2.4 $\pm$ 0.1 & 9 $\pm$ 4 & 0.8 $\pm$ 0.3 & 2.34 $\pm$ 0.10 \\
\SI{1.49}{pW} & 0.481 $\pm$ 0.007 & 1.73 $\pm$ 0.05 & 14 $\pm$ 3 & 1.2 $\pm$ 0.3 & 1.80 $\pm$ 0.05 \\
\SI{556}{fW} & 0.309 $\pm$ 0.005 & 1.11 $\pm$ 0.04 & 13 $\pm$ 3 & 1.0 $\pm$ 0.3 & 1.70 $\pm$ 0.06 \\
\SI{147}{fW} & 0.229 $\pm$ 0.002 & 0.98 $\pm$ 0.03 & 21 $\pm$ 2 & 1.2 $\pm$ 0.2 & 1.29 $\pm$ 0.03 \\
\SI{37.6}{fW} & 0.201 $\pm$ 0.002 & 0.69 $\pm$ 0.09 & 60 $\pm$ 20 & 0.7 $\pm$ 0.1 & 1.24 $\pm$ 0.07 \\
\SI{8.42}{fW} & 0.247 $\pm$ 0.002 & 0.77 $\pm$ 0.08 & 60 $\pm$ 20 & 0.7 $\pm$ 0.1 & 1.14 $\pm$ 0.05 \\
\SI{2.79}{fW} & 0.213 $\pm$ 0.002 & 0.89 $\pm$ 0.05 & 34 $\pm$ 5 & 1.0 $\pm$ 0.2 & 1.06 $\pm$ 0.04 \\
\bottomrule
\end{tabular}
\label{tab:bbfitparams}
\end{table}

\begin{table}[tbp]
\centering
\caption
[Best-fit parameters from the spectral density fits of the continuous-wave data.]
{Continuous-wave best-fit parameters with uncertainties.}
\renewcommand{\arraystretch}{1.2}
\begin{tabular}{cccccc}
\toprule
$P_A$ & $A^2 \; [\SI{e-18}{Hz^{-1}}]$ & $W^2 \; [\SI{e-18}{Hz^{-1}}]$ & $f_\knee \; [\si{Hz}]$ & $\spectralindex$ & $f_\cutoff \; [\si{kHz}]$ \\
\midrule
\SI{29.0}{pW} & 4.89 $\pm$ 0.06 & 1.3 $\pm$ 0.1 & 330 $\pm$ 40 & 1.46 $\pm$ 0.05 & 2.5 $\pm$ 0.4 \\
\SI{18.1}{pW} & 3.20 $\pm$ 0.04 & 1.46 $\pm$ 0.09 & 270 $\pm$ 30 & 1.33 $\pm$ 0.05 & 2.6 $\pm$ 0.3 \\
\SI{9.72}{pW} & 1.82 $\pm$ 0.03 & 1.27 $\pm$ 0.07 & 220 $\pm$ 30 & 1.28 $\pm$ 0.06 & 2.4 $\pm$ 0.2 \\
\SI{4.89}{pW} & 0.98 $\pm$ 0.01 & 1.22 $\pm$ 0.05 & 160 $\pm$ 20 & 1.25 $\pm$ 0.06 & 2.2 $\pm$ 0.1 \\
\SI{1.93}{pW} & 0.462 $\pm$ 0.006 & 0.97 $\pm$ 0.04 & 100 $\pm$ 10 & 1.09 $\pm$ 0.07 & 1.83 $\pm$ 0.07 \\
\SI{573}{fW} & 0.288 $\pm$ 0.003 & 0.87 $\pm$ 0.04 & 52 $\pm$ 7 & 1.1 $\pm$ 0.1 & 1.56 $\pm$ 0.05 \\
\SI{176}{fW} & 0.244 $\pm$ 0.003 & 0.88 $\pm$ 0.06 & 37 $\pm$ 8 & 0.9 $\pm$ 0.2 & 1.29 $\pm$ 0.05 \\
\SI{48.7}{fW} & 0.219 $\pm$ 0.002 & 0.82 $\pm$ 0.05 & 39 $\pm$ 7 & 1.0 $\pm$ 0.1 & 1.21 $\pm$ 0.05 \\
\SI{13.0}{fW} & 0.210 $\pm$ 0.002 & 0.7 $\pm$ 0.2 & 40 $\pm$ 40 & 0.6 $\pm$ 0.2 & 1.17 $\pm$ 0.08 \\
\SI{2.09}{fW} & 0.235 $\pm$ 0.002 & 0.85 $\pm$ 0.08 & 40 $\pm$ 10 & 0.8 $\pm$ 0.2 & 1.13 $\pm$ 0.05 \\
\bottomrule
\end{tabular}
\label{tab:cwfitparams}
\end{table}

\end{document}